# Role of the orbital degree of freedom in iron-based superconductors

Ming Yi[1,2†], Yan Zhang[1,3†], Zhi-Xun Shen[1,2*], and Donghui Lu[4*]

[1]*Stanford Institute for Materials and Energy Sciences, SLAC National Accelerator Laboratory and Stanford University, Menlo Park, California 94025, USA*

[2]*Departments of Physics and Applied Physics, and Geballe Laboratory for Advanced Materials, Stanford University, Stanford, California 94305, USA*

[3]*Advanced Light Source, Lawrence Berkeley National Lab, Berkeley, California 94720, USA*

[4]*Stanford Synchrotron Radiation Lightsource, SLAC National Accelerator Laboratory, Menlo Park, California 94025, USA*

[†]*These authors contributed equally*

*To whom correspondence should be addressed: dhlu@slac.stanford.edu and

zxshen@stanford.edu

**Abstract**

Almost a decade has passed since the serendipitous discovery of the iron-based high temperature superconductors (FeSCs) in 2008. The fact that, as in the copper oxide high temperature superconductors, long-range antiferromagnetism in the FeSCs arises in proximity to superconductivity immediately raised the question of the degree of similarity between the two. Despite the great resemblance in their phase diagrams, there exist important differences between the FeSCs and the cuprates that need to be considered in order to paint a full picture of these two families of high temperature superconductors. One of the key differences is the multi-orbital multi-band nature of the FeSCs, which contrasts with the effective single-band nature of the cuprates. Systematic studies of orbital related phenomena in FeSCs have been largely lacking. In this review, we summarize angle-resolved photoemission spectroscopy (ARPES) measurements across various FeSC families that have been reported in literature, focusing on the systematic trends of orbital dependent electron correlations and the role of different Fe 3d orbitals in driving the nematic transition, the spin-density-wave transition, and superconductivity.

**Introduction**

Thirty years after the historic discovery of cuprate high temperature superconductors, the mechanism for high temperature superconductivity remains the biggest challenge in condensed matter physics despite tremendous amount of theoretical and experimental efforts. The discovery of iron-based superconductors[1] provides a great opportunity to identify the important ingredients that are common to both families of high Tc materials and to test the theoretical models that have been formulated for cuprates. Comparing FeSCs with cuprates, the most striking similarity is the common phase diagram, in which unconventional superconductivity appears in the vicinity of other competing phases, such as the pseudogap phase and the charge order in cuprates and the spin-density-wave (SDW) phase and nematic phase in FeSCs[2-3]. The emergence of superconductivity always takes place with the suppression of these competing phases. Such a remarkable resemblance has raised our hope for a unified theory of high temperature superconductivity and has motivated many theorists to take the strong coupling approach to describe FeSCs.

On the other hand, FeSCs also appear to distinguish themselves from cuprates in various aspects, including metallicity of the parent phase, crystal structure of the conduction layer, spin symmetry of the antiferromagnetic order, as well as the underlying electronic structure. Prior to establishing a unified understanding of the physics in cuprates and FeSCs, we first need to understand whether these differences are trivial nuances or critical ingredients that cannot be neglected. Among them, the most fundamental difference is the multi-orbital multi-band nature of the underlying electronic structure in FeSCs. In contrast to cuprates, for which essential physics seems to take place in a single effective band and Fermi surface, there are at least three out of five Fe 3d orbitals that are active near the Fermi level ($E_F$) in FeSCs, forming multiple Fermi surface sheets. The complexity of theoretical treatment for a multi-band system has led to various proposals for minimal models for FeSCs, in which the orbital degree of freedom is often ignored for simplicity. While these models capture some underlying physics, the question is whether they miss important orbital related physics.

The lack of systematic experimental studies on the role of different Fe 3d orbitals may be part of the reason that orbital physics in FeSCs has not garnered as much attention as they perhaps deserved. In this review, we summarize experimental evidence of various orbital dependent phenomena in the electronic structure from angle-resolved photoemission spectroscopy literature. For a recent review on the progress of theoretical studies on the orbital degree of freedom, please see Ref. 4. First, we discuss systematic trends on the normal state electronic structure across different families of FeSCs. In particular, we point out a systematic change of the orbital related physics, for example the increase of orbital-selective electron correlations across the wide spectrum of FeSC compounds based on both structural parameters and charge carrier doping. Eventually, this growth of orbital-dependent correlations naturally culminates in a complete localization of a single orbital while other orbitals remain itinerant—an orbital-selective Mott phase. Hence the orbital degree of freedom plays an important role in balancing the coexistence of local and itinerant physics in the normal state of the FeSCs. We then focus on the manifestations of different Fe 3d orbitals across the nematic and the SDW transitions. In the nematic state, strong anisotropy occurs in the hopping of the $d_{xz}$, $d_{yz}$, and $d_{xy}$ orbitals as characterized by a momentum dependent reconstruction of $d_{xz}$, $d_{yz}$, and $d_{xy}$ bands. In the SDW state, bands fold and SDW gaps open. The SDW gap is also strongly orbital-dependent. It is large on the $d_{xy}$ bands but moderate on the $d_{xz}/d_{yz}$ bands. These observations have strong implications on the theories aiming to understand the nature of those competing phases and suggest an important role of multi-orbital nature of FeSCs in driving the nematic and SDW transitions. In the end, we discuss the non-trivial implications of multi-orbital nature on the superconducting pairing mechanism of these materials.

**The normal state**

All FeSC compounds, regardless of their compositions, share in common planes containing iron and pnictogen (P/As) or chalcogen (S/Se/Te), which is located alternatingly above and below the Fe lattice (Fig. 1a). The difference among FeSC families is the composition and structure of interlayers between the Fe planes. In some cases, these interlayers form charge reservoirs that donate charge carriers to the Fe planes. The normal state is defined as the phase outside the boundaries of the magnetic and structural transitions (Fig. 1a). All FeSC compounds share in common the basic electronic structure consisting of Fe 3d bands near $E_F$, with the $d_{xz}$, $d_{yz}$, and $d_{xy}$ orbitals most active near $E_F$. The basic electronic structure of FeSC in the normal state is illustrated in Fig. 1c, where three hole-like bands reside near the Brillouin Zone (BZ) center, $\Gamma$, and two electron-like bands near the BZ corner, M (except in the case of the 122 structure, where this point is called the X point). For different doping levels and structural subtleties, the overall or relative positions of the hole and electrons bands may vary in energy, leading to different Fermi surface topologies with varying number of hole pockets at $\Gamma$ and electron pockets at M (Fig. 1d). For undoped parent compounds, the hole pockets at $\Gamma$ and electron pockets at M are of similar sizes. For hole-doped compounds, the hole pockets at $\Gamma$ enlarge while the electron pockets shrink. For heavily electron-doped compounds such as the electron-doped chalcogenides, the hole pockets disappear while large electron pockets remain at M.

One of the fundamental questions after the discovery of the FeSCs was whether it is appropriate to model them as localized systems or itinerant systems. On one hand, the observed large spectral weight in the fluctuating magnetic spectrum[5] tends to suggest the former, while the high density of states found near $E_F$ compared to the cuprates[6] seems to suggest the latter. As we know now, neither picture is fully complete. We will show in the following that there is a large systematic spread of electron correlation strength among different FeSCs, and more importantly, this occurs in a strongly orbital-dependent way.

This trend can be qualitatively seen by comparing representative compounds from the iron phosphides to the iron arsenides and to the iron chalcogenides (Fig. 2). Electron correlation renormalizes the electronic bandwidth. From ARPES data, one way to quantify the strength of correlation is to extract the ratio of the non-interacting bandwidth calculated from local density approximation (LDA) and experimentally measured bandwidth, which is the renormalization factor[7]. Fig. 2a-c shows the band dispersions along the high symmetry direction Γ-M observed in the paramagnetic state of a phosphide ($SrFe_2P_2$), an arsenide (NaFeAs), and a chalcogenide ($FeSe_{0.44}Te_{0.56}$), plotted in the same energy window, where the $d_{yz}$ band is marked in green and $d_{xy}$ blue. There are two observations to make: i) the bandwidth of all bands systematically narrows from phosphide to the arsenide to the chalcogenide, which is also seen in the increase of the renormalization factors, and ii) the bandwidth of $d_{xy}$ narrows at a much faster rate than that of $d_{yz}$ from the phosphides to the chalcogenides. These two observations suggest that both the overall electron correlation strength and orbital-selectivity, which is defined here as the ratio in the band renormalization between the $d_{xy}$ band and $d_{yz}$ band, increase from the iron phosphides to the iron chalcogenides. In the following section, we systematically demonstrate these trends and discuss factors that correlate with these two parameters.

Considering the three active orbitals near $E_F$, the $d_{xz}$ and $d_{yz}$ orbitals are bound by $C_4$ symmetry to be degenerate in the tetragonal state. The $d_{xy}$ orbital, however, does not necessarily need to behave in the same way as $d_{xz}/d_{yz}$. Hence we examine each orbital in turn. For the bands near $E_F$, the $d_{yz}$ band is the band that can be observed most completely under common polarization setups, with its band top at $\Gamma$ and band bottom between $\Gamma$ and M (Fig. 1d). Here we extract its bandwidth from ARPES literature and compare with available LDA calculations (Table 1). Figure 2d shows the $d_{yz}$ renormalization factors for FeSC compounds measured in the normal state as a function of the Fe-Pn/Ch bond length. To avoid complications from doping effects, we only plot undoped parent compounds here where the electron filling, n, is 6. First of all, we see a clear trend with a continuous spread of renormalization factors from the phosphides at the bottom to the 122 arsenides to the 111 arsenides and to the chalcogenides. The $d_{yz}$ renormalization factor, which is a measure of overall correlation strength, increases with the Fe-Pn/Ch bond length. Intuitively, longer bond length reduces electron hopping and therefore the kinetic energy, hence leading to more localized behavior. However, similar plot for the $d_{yz}$ and $d_{xy}$ renormalizations as a function of the lattice $a$ (effectively Fe-Fe bond length) does not show as clear a trend over all compounds (Fig. 3a,c). As $d_{yz}$ is extended in the out-of-plane direction and $d_{xy}$ is mostly in plane, the lack of clear trend in both renormalizations against lattice $a$ suggests that electron hopping is dominated by the indirect path of Fe-Pn/Ch-Fe rather than that of direct Fe-Fe bond. The $d_{yz}$ renormalization factor also has strong correlation with related structural parameters such as anion height and bond angle, which both directly relate to the Fe-Pn/Ch bond length[8]. Similar dependence of correlation strength on structural parameters is also seen in optical data within the $BaFe_2As_2$ family[9] and photoemission data within the heavily electron-doped iron chalcogenides[10]. For other structural parameters such as the lattice constant $c$, the correlation with bandwidth renormalization is not obvious across families (Fig. 3b,d).

Structural parameters are not the only factors that correlate with the overall correlation strength. A second factor is electron filling[7]. In Fig. 2e we plot the $d_{yz}$ bandwidth renormalization versus the electron filling for doped compounds of two series, the $BaFe_2As_2$ series and the $LiFeAs$ series. To put this plot in perspective, we also overlay the electron (Co) and hole (K) doped phase diagram of $BaFe_2As_2$ on the horizontal axis. The data points for this series range from those taken from $KFe_2As_2$ (n = 5.5) to $BaNi_2As_2$ (n = 8). Here we note that the undoped parent compounds of FeSC has n = 6, which is not half-filling as the case of the parent compounds of the cuprates. True half-filling for the Fe 3d orbitals is n = 5. This explains the asymmetry of the overall correlation with respect to the undoped parent compounds of FeSC[7,9]. The electron correlation is weak far away from n = 5 for the heavily electron-doped compounds, and diverges towards n = 5 on the hole-doped side. However, it is interesting to note that, under this scenario, the known undoped iron pnictides are effectively on the electron-doped side of the true half filling. In analogy to the cuprates, there may be an equivalent regime of superconductivity on the hole-doped side, as has been recently theoretically suggested[11-12].

In Fig. 2f we plot all the compounds (including those with different electron fillings) with sufficient information from literature on a 2D plot with electron filling and the Fe-Pn/Ch bond length, and use color to indicate the strength of $d_{yz}$ bandwidth renormalization factor. Here we see as demonstrated before, both reducing electron filling towards n = 5 and lengthening the Fe-Pn/Ch bond length increase overall correlation.

Next, we discuss factors that affect the orbital-selectivity. Since the $d_{xy}$ orbital is the most correlated of the orbitals, quantitatively, we extract the ratio of the renormalization factors of the $d_{xy}$ band and the $d_{yz}$ band. In Fig. 2h we plot this for all compounds we found from literature against the Pn/Ch-Fe-Pn/Ch bond angle. A clear trend is seen where smaller bond angle, as in the case of the chalcogenides, leads to strong selectivity; whereas bigger bond angle, as in the case of the phosphides, leads to almost no selectivity. This is because smaller bond angle results in a vertically elongated tetragon, which reduces hopping more dramatically for the in-plane $d_{xy}$ orbital than $d_{xz}/d_{yz}$ orbitals, considering hopping is dominantly mediated through the Pn/Ch, which reside out of the Fe plane. Besides the bond angle, Fig. 2g also shows another factor that correlates with orbital-selective correlation—the overall electron correlation represented by the $d_{yz}$ renormalization factor. It demonstrates that among all FeSCs, the stronger the overall correlation, the stronger the orbital-selectivity in the $d_{xy}$ orbital. This has been discussed in previous theoretical work as due to the tendency towards a Hund's metal phase, where Hund's coupling increases orbital differentiation and independence by suppressing inter-orbital correlations[8,13-14]. In Fig. 2i, we plot the orbital-selectivity as a function of both bond angle and overall correlation strength for all compounds. We see that these dependences are not only true within a family of FeSC, but systematically spread among all FeSC compounds.

As orbital-selectivity increases, an interesting phenomenon occurs in the strongly correlated members of FeSCs—a tendency towards an orbital-selective Mott phase (OSMP). In Fig. 4a, we plot the bond length dependence of the renormalizaton factors of the $d_{xy}$ orbital for undoped FeSCs (n = 6), which is the most correlated orbital. Comparing to the equivalent plot for $d_{yz}$ (Fig. 2d), we see that the dynamic range of the $d_{xy}$ orbital is five-fold of that of $d_{yz}$. This again showcases the strong orbital differentiation among the FeSCs towards the strongly correlated members. When this differentiation is strong, as in the iron chalcogenides, the normal state of these materials is sufficiently close to an OSMP such that raising temperature has been observed to push them into the OSMP where the $d_{xy}$ orbital completely loses its spectral weight while other orbitals remain itinerant (Fig. 4b)[15]. This has in fact been observed universally for different iron chalcogenides including Fe(Te,Se), $KFe_2Se_2$, and monolayer FeSe film grown on $SrTiO_3$[16-19] and $Nb:BaTiO_3/KTaO_3$ heterostructures[20]. Even for the most correlated iron arsenide, $KFe_2As_2$, evidence for decoherence of the $d_{xy}$ orbital has been observed by transport measurements[21]. Here we see that the tendency towards the OSMP in the chalcogenides is not accidental, but grows naturally from an increasingly selective orbital correlation systematically in all FeSCs.

Both the spread of electron correlation strength and the increasing orbital selectivity are perhaps what makes it difficult to develop a unified theoretical model for all FeSCs. On one side of the spectrum the orbitals are mostly itinerant with almost no orbital selectivity while on the other side strong orbital decoupling due to Hund's coupling results in one of the orbitals approaching localization. Nonetheless, this orbital-selectivity is a behavior unique and essential to the multi-orbital physics of the FeSCs in contrast to the single-band physics of the cuprates. Its importance is especially evident for the compounds on the strongly correlated side such as the chalcogenides. These behaviors have led to interesting theoretical proposals where one redefines the phase diagram using an average orbital filling, showing a gradual approach from superconductivity towards a Mott insulating state via an OSMP[13], while another proposes that in the strongly orbital-selective regime, the $d_{xy}$ orbital is near half-filling[14]. Both suggest parallels reminiscent of the cuprate problem.

**<u>The nematic state</u>**

In a typical phase diagram of iron-based superconductors, the material in the underdoped regime goes through two transitions and enters the nematic state and the SDW state at low temperatures[2-3]. The multi-orbital nature not only leads to orbital-dependent band renormalization in the normal state, but also plays an important role in driving the system into these symmetry-breaking states. The nematic phase is marked by a tetragonal to orthorhombic structural transition at $T_S$, where $C_4$ rotational symmetry is broken down to $C_2$ symmetry without breaking the translational symmetry. With the orthorhombic distortion, material forms natural twin domains. Hence for macroscopic probe like ARPES where the beam spot is larger than the typical domain size (Fig. 5a), the photoemission signal becomes the superposition of band structures from two orthogonal domains. As a result, complex band reconstructions were reported in early ARPES studies[22-24]. Subsequently, it was realized that samples with a single domain could be achieved by applying a uniaxial pressure mechanically on the sample[25-26]. The intrinsic band structure of the $C_2$ state was then revealed by ARPES measurements on detwinned crystals of $BaFe_2As_2$[27-29], $NaFeAs$[30-31], $FeSe$[32-35], and FeSe films[36-37]. Figure 5b shows an example of detwinned NaFeAs measured in the nematic state, where the $d_{yz}$ band (green) along the slightly longer axis is shifted up while the $d_{xz}$ band (red) along the shorter axis is shifted down. In the tetragonal state above $T_S$, these two bands are degenerate in energy. This shift in opposite directions in the orthorhombic state indicates the breaking of the $C_4$ symmetry in the orbital degree of freedom. This generic upward shift of $d_{yz}$ orbital and downward shift of $d_{xz}$ orbital is observed in all measurements of detwinned FeSCs in the nematic phase[27-37]. Importantly, the relatively large energy scale of this orbital splitting cannot be a trivial consequence of the less than 1% orthorhombicity[27]. Hence the orbital anisotropy is unlikely to be a simple result of the lattice, but rather suggests the manifestation of electronic nematicity, consistent with the original discovery of large resistivity anisotropy in this phase[25]. We note that, on top of the discussed band separation, an additional small band splitting has been observed by ARPES in FeSe[38-39], which induced hot debates on how to extract the correct nematic energy splitting from the ARPES data. The energy scale of the large band separation scales with the nematic transition temperature in all measured FeSCs and thus likely represents the strength of nematic order. Moreover, very recent data on detwinned FeSe clearly shows the two bands with large energy separation to belong to orthogonal directions[40], hence consistent with the previous understanding that the $d_{xz}$ and $d_{yz}$ orbitals are indeed split by ~50meV in the nematic phase, similar to the pnictides.

The discovery of energy splitting of $d_{xz}$ and $d_{yz}$ bands has been viewed as an evidence for the existence of orbital order in the nematic state. It has been proposed that the on-site occupation difference between $d_{xz}$ and $d_{yz}$ orbitals triggers a ferro-orbital ordering, which consequently drives the system into the nematic state[41-43]. In most band calculations where only the on-site occupation difference is considered, the ferro-orbital order results in an energy splitting of $d_{xz}$ and $d_{yz}$ bands that is almost constant throughout the BZ. Experimentally, clear delineation of the momentum dependence of the $d_{xz}$ and $d_{yz}$ splitting has not been carefully examined for most FeSCs, because the close proximity of the magnetic and structural transitions prevents a clear separation of the effects of the two phases. FeSe, which does not have a magnetic transition[44], offers a clean case for examining the nature of nematic order and associated nematic transition in detail. Figure 5c shows the band dispersions along the high symmetry direction Γ-M of a twinned multi-layer FeSe film grown on $SrTiO_3$[36], which resembles bulk FeSe crystal. On a twinned sample, the longer and shorter orthogonal high symmetry directions from orthogonal domains are superpositioned along the same cut. Hence the shifted $d_{xz}/d_{yz}$ bands appear as effectively split in energy below $T_S$. From this data, two important observations can be made. Firstly, the $d_{xz}/d_{yz}$ bands, which are degenerate above $T_S$, are split in the same fashion as in other FeSCs. Secondly, a clear momentum-dependence is seen in this orbital anisotropy. As shown in Fig. 5d, instead of being constant, a finite energy splitting at $\Gamma$ first decreases to zero and then increases again, reaching its maximum at the M point. The momentum-dependence of the energy splitting is consistent with $BaFe_2As_2$ and $NaFeAs$, where the orbital anisotropy is small at $\Gamma$ but large at M[27]. Furthermore, the momentum dependence of orbital anisotropy is non-monotonic, which may be understood by considering a band splitting ($E_{yz}-E_{xz}$) that switches sign between $\Gamma$ and M (Fig. 5e). This has indeed been experimentally observed in detwinned FeSe bulk crystal[35,40]. This non-trivial momentum dependence of the $d_{xz}/d_{yz}$ energy splitting suggests that the simple on-site occupation difference between $d_{xz}/d_{yz}$ orbitals is unlikely to be the driving force of the nematic phase. Instead, other mechanisms such as anisotropic hopping of the $d_{xz}/d_{yz}$ orbitals[45-46], Pomeranchuk instability[47], orbital-dependent correlation effect[48], and orbital-selective spin fluctuations[49] could all result in different renormalization and/or shift of the $d_{xz}/d_{yz}$ bands in the nematic state.

While the $d_{xz}/d_{yz}$ orbital anisotropy has attracted great attentions, the $d_{xy}$ orbital has not been considered in most theoretical models, partly due to the lack of experimental evidence for a clear participation of $d_{xy}$ orbital in the electronic reconstructions and the assumption that it does not contribute to $C_4$ symmetry breaking. However, as illustrated in Figs. 5f and 5g, to fully explain the three electron bands observed in the nematic phase at the M point on this twinned sample, the shift of the $d_{yz}$ and $d_{xz}$ bands in opposite directions is insufficient, in addition, the $d_{xy}$ band along the longer x-direction must shift down in energy while the counterpart $d_{xy}$ band along the shorter y-direction must remain unshifted[36]. Furthermore, the magnitude of the energy splitting of the $d_{xy}$ band along two perpendicular directions is comparable to the energy splitting of the $d_{xz}/d_{yz}$ orbitals, which further suggests the complexity of the orbital anisotropy in the nematic state.

The observed band reconstruction and its orbital dependence put strong constraints on theoretical models regarding the nematic state. On one hand, while most theories consider only the $d_{xz}/d_{yz}$ orbitals, the ARPES results suggest that all three orbitals play an important role in driving the nematic state. As one theoretical study shows, the coupling of the $d_{xy}$ orbital to the nematic order parameter is necessary for predicting the correct gap symmetry in FeSe[50]. On the other hand, the energy scale of the band reconstruction shows a strong momentum dependence, indicating that the anisotropy is unlikely to be dominated by the on-site occupation difference between the $d_{xz}$ and $d_{yz}$ orbitals. Instead, other mechanisms must be considered to correctly describe the momentum-dependence orbital anisotropy.

**The spin-density wave order**

Next, we discuss the role of different orbitals in forming the collinear SDW order. The SDW order has been found to couple strongly with the nematic order in most cases. On top of the rotational symmetry breaking, the SDW order further breaks the translational symmetry. As a result, the BZ reduces and bands fold across the SDW zone boundary. Spectrally, the signatures of the SDW band reconstruction are very distinct from that of the nematic phase. Instead of the band shift, SDW gaps open where folded bands cross original bands. It is simpler to illustrate the details of the band folding by unfolding the bands into the 1-Fe BZ, where the folding occurs strictly along the AFM direction $\Gamma$-$M_x$ but not along the FM direction (Fig. 6a). Here, the $\Gamma$ point has two hole pockets from the $d_{xz}$ and $d_{yz}$ hole bands. At the $M_x$ point, an electron pocket appears that is of $d_{xy}$ character along the AFM direction and $d_{yz}$ along the FM direction. At the $M_y$ point, another electron pocket appears with $d_{xy}$ along the FM direction and $d_{xz}$ along the AFM direction. At the $\Gamma'$ point, in some of the iron-arsenide compounds, lives the third hole pocket of $d_{xy}$ character. With SDW folding, the $\Gamma$ and $M_x$ points fold unto each other, while the $M_y$ and $\Gamma'$ points fold unto each other.

Next, we discuss the detailed band reconstruction around these two folded points using data from detwinned $BaFe_2As_2$, which can be revealed by using different polarizations. For the folded $\Gamma$-$M_x$ cut (Fig. 6b), along the AFM direction, the $d_{xy}$ electron band from $M_x$ folds unto the $d_{xz}$ and $d_{yz}$ hole bands around $\Gamma$, but little or no SDW gap appears along this high symmetry direction where folded bands cross. Along the orthogonal FM direction at this same point (Fig. 6c), the $d_{yz}$ electron band crosses the $d_{xz}$/$d_{yz}$ bands, forming an SDW gap on the order of ~30meV between the $d_{yz}$ bands, which can be seen in both the $d_{yz}$ hole band and the $d_{yz}$ electron band. For the folded $M_y$-$\Gamma'$ point (Fig. 6d), along the AFM direction, the $d_{xz}$ electron band crosses the $d_{xy}$ hole band, opening an SDW gap that saddles $E_F$, with a gap size bigger than 50meV. Along the orthogonal FM direction at this point (Fig. 6e), the $d_{xy}$ hole and electron bands cross each other, opening an SDW gap that is larger than 50meV. Overall, we see that the SDW gap opens at the folded band crossings in an orbital-dependent way. The biggest SDW gaps occur in the $d_{xy}$ bands around the $M_y$-$\Gamma$' point. The SDW gap is moderate for the $d_{yz}$ bands at the $\Gamma$-$M_x$. For the $d_{xz}$ hole bands, the SDW gap is the smallest, resulting in dominant $d_{xz}$ orbital weight near the Fermi energy in the SDW state[51]. Four SDW gap nodes exist along the high symmetric directions due to the incompatible symmetries of the crossing bands[52]. As a result, the Fermi surface reconstructs drastically, forming four small Fermi pockets along both the $\Gamma$-$M_x$ and $\Gamma$-$M_y$ directions (lower panel of Fig. 6a).

The SDW order has been described in both weak- and strong- coupling theories. It is still largely debated whether the SDW is originated from the super-exchange interaction of local moments or the Fermi surface nesting of itinerant electrons[5,53-58]. The orbital-dependence of the band reconstruction in the SDW state suggests an orbital-selective magnetism in the iron-based superconductors. It has been proposed that the magnetic moments originate mainly from the $d_{xy}$ and $d_{yz}$ orbitals instead of the $d_{xz}$ orbital in the SDW state[59-61], which is consistent with the orbital-dependent SDW gaps observed here. The multi-orbital nature of the SDW state also explains the dual character of the magnetic moments observed by neutron scattering experiment[53]. Both the local property of $d_{xy}$ orbital and the itinerancy of the $d_{yz}/d_{xz}$ orbitals should be considered in constructing the microscopic model of the SDW state in iron-based superconductors.

**The superconducting state**

Finally, we discuss the role of orbitals in the superconducting state. While the superconducting gap in cuprates can be fitted by a simple $d_{x2-y2}$ gap function across all different families, the gap symmetry in FeSCs has a complex distribution in momentum space that varies among different families[62]: while largely isotropic gaps have been observed in $Ba_{1-x}K_xFe_2As_2$[62] consistent with s± symmetry, strong gap anisotropy or even gap nodes have been reported in $BaFe_2(As_{1-x}P_x)_2$[63-64], $LiFeP$[65], bulk $FeSe$[66], $FeSe_{1-x}S_x$[67], and FeSe films[68]. In cases like 1ML FeSe thin film, the gap function cannot be fitted by single trigonometric gap functions under s±, d-wave, or extended s-wave gap symmetries[69]. This contrasting behavior is intimately connected to the multi-band multi-orbital nature of the underlying electronic structure and again implies the non-trivial role which orbital physics plays in the FeSCs.

Frist of all, the Fermi surface of FeSCs consists of multiple Fermi pockets with distinctive orbital characters. Such Fermi surface topology allows Cooper pairs to scatter via intra- or inter-orbital scattering channels. One critical question is which scattering channel is more important for superconducting pairing. The intra-orbital pairing would result in an orbital dependent superconducting gap while the inter-orbital pairing tends to generate equivalent gaps among different orbitals. Experimentally, multi-gap behavior has been observed in several iron-based compounds by scanning tunneling microscopy (STM)[70-71] and transport measurements[72-73]. ARPES studies on $Ba_{1-x}K_xFe_2As_2$ and LiFeAs show that the superconducting gap on the $d_{xz}/d_{yz}$ hole pockets is much larger than that on the $d_{xy}$ hole pocket[62]. All these results seem to suggest the dominating role of intra-orbital pairing in FeSCs.

Another important question often asked is which orbital plays the most important role for superconductivity. This, too, seems to vary between different compounds. For most iron-pnictide compounds studied with comparable hole and electron pockets, the $d_{xz}/d_{yz}$ hole pockets show larger superconducting gap than the $d_{xy}$ hole pocket[62,74], suggesting the importance of $d_{xz}/d_{yz}$ orbitals to superconductivity. Consistently, in $LiFe_{1-x}Co_xAs$ and $Ba(Fe_{1-x}Co_x)_2As_2$, when the $d_{xz}/d_{yz}$ hole pockets vanish with electron doping, superconductivity is strongly suppressed[75-76]. On the contrary, for heavily electron-doped iron-selenide superconductors with only electron pockets, the superconducting gap of 1ML FeSe film show maxima on the $d_{xy}$ electron bands[77]. Furthermore, in bulk FeSe crystal, where the sample is continuously surface-doped to enhance superconductivity, it is reported that the appearance of the $d_{xy}$ electron pocket on the Fermi surface at the M point coincides with the beginning of the second superconducting dome with higher $T_C$ and different pairing symmetry[77]. Interestingly, another work on the pnictide compound $Ca_{10}(Pt_4As_8)(Fe_{2-x}Pt_xAs_2)_5$ shows relatively high $T_C$ with the presence of only the $d_{xy}$ hole pocket at $\Gamma$[78]. This may on one hand suggest the importance of the $d_{xy}$ orbital to superconductivity, on the other hand highlight the strong correlation between inter-pocket (intra-orbital) scattering and superconductivity.

The level of material-dependence reported in the past decade has been somewhat puzzling and perhaps disappointing for the ultimate goal of finding a simple unifying description of superconductivity. However, this may be well expected when we consider the multi-orbital nature of the FeSCs. As this is a new dimension which has been lacking from the machinery developed out of the cuprate problem, theoretical work taking into account the orbital degree of freedom has been very limited, but several work have already showed promise. From the strong coupling approach using a multiorbital $t$-$J_1$-$J_2$ model, one study showed that the orbital-selectivity results in a gap anisotropy that is also orbital-dependent[79]. From a weak-coupling approach, a very recent theoretical study based on spin fluctuations taking into consideration the orbital-selective renormalization that modulates the coherent spectral weight of different orbitals was able to reproduce the observed momentum-dependent gap structure of monolayer FeSe and LiFeAs[80]. Other theoretical works have also proposed interesting mechanisms by which the FeSCs and cuprates could be united[81-82]. These work importantly demonstrate that behind the apparent gap variations amongst FeSCs there may be a common underlying pairing mechanism, and the source of the material-dependence may be the different degree of orbital-selective correlation effects, which tune the dominant orbitals that are manifested.

**Discussion**

We first summarize the key findings of the four major phases discussed:

- The normal state:
  - Electron correlations systematically and continuously spread across FeSC families tuned by two factors: i) the Fe-Pn/Ch bond length, and ii) electron doping away from true half-filling at n=5.
  - Electron correlations are orbital-dependent, with the $d_{xy}$ orbital being the most localized. The strength of orbital-selectivity is tuned by i) overall correlation strength, and ii) bond angle, eventually reaching an orbital-selective Mott phase where the $d_{xy}$ orbital is completely localized while other orbitals remain itinerant.
- The nematic state:
  - There is significant $C_4$ symmetry breaking in the orbital degree of freedom at the onset of the structural transition beyond the effect of the lattice distortion. In particular, the $d_{yz}$-dominant band is observed to shift up while the $d_{xz}$-dominated band is observed to shift down, albeit in a strongly-momentum dependent way.
  - The $d_{xy}$ orbital is also observed to participate by exhibiting a splitting in energy that is comparable to that of $d_{xz}/d_{yz}$ orbitals, suggesting an anisotropic hopping origin rather than ferro-orbital order.
- The spin density wave order:
  - Band folding occurs, producing SDW gaps that are the largest in the $d_{xy}$ orbital, moderate in the $d_{yz}$ orbital, and smallest in the $d_{xz}$ orbital.
- The superconducting state:
  - Superconducting gaps are generally observed to be multi-gap, suggesting

dominance of intra-orbital pairing.

- Gap functions cannot be described by single trigonometric gap functions, and also vary among families, suggesting the complex role of intra-orbital pairing and multi-orbital FS.

Having discussed the normal state, the nematic state, the magnetic state, and the superconducting state separately, we now discuss the relationship between these phases. From the normal state properties, we see that there is a systematic spread of electronic correlation over all the FeSCs, with a large dependence on certain structural parameters such as the bond length and bond angle. As has been shown, the superconducting temperature, $T_C$, is also highly dependent on the bond angle[83]. Hence superconductivity is expected to be optimized at intermediate electron correlation strength. The nematic phase and the collinear SDW phase are often discussed together. Here we see that the two orders can be strongly coupled, as in most iron arsenides, but not necessarily always the case, as in FeSe. Regardless of the strength of coupling of these two orders, we see that the spectral signature and magnitude for the nematic order is the same across different materials. The nature of these two phases to superconductivity is competitive. As has been reported, the spectral order parameters of these two orders both decrease at the onset of superconductivity[84], similar to the macroscopic order parameters of the lattice orthorhombicity and the magnetic moment[85].

For all three phases discussed, we also see a strong orbital-dependence. Hund's coupling suppresses orbital interaction, separating the $d_{xy}$ orbital from the largely degenerate $d_{xz}/d_{yz}$. This coupled with crystal field splitting effectively makes the $d_{xy}$ orbital the most strongly correlated, as seen in the normal state, and in some cases close to half-filling while the overall filling is effectively on the electron-doped side. This occurrence enables some behaviors of the $d_{xy}$ orbital in the FeSCs to become relevant to the cuprates, such as the observation of OSMP in the most correlated iron chalcogenides. We have also seen that the $d_{xy}$ orbital actively participates in the nematic and magnetic competing phases to superconductivity, by both developing a non-trivial anisotropy below $T_S$ and the largest SDW gap below $T_{SDW}$. The $d_{xz}/d_{yz}$ orbitals, on the other hand, are less correlated than $d_{xy}$ and maintains a certain level of itinerancy due to their degeneracy and effective electron-doping away from half-filling. This degeneracy also gives them a more active role in the nematic phase, where the degeneracy is lifted below $T_S$.

Overall, we see that the orbital-dependence in the FeSC plays a nontrivial role. The systematic spread of orbital-selective electron correlations across different FeSC families may well be the origin of material-dependent variations that is manifested when the participation of different orbitals is enhanced or suppressed in different phases. When taken under this perspective, there is still hope that the underlying mechanism for superconductivity and competing phases in the FeSCs and perhaps even the cuprates may take on a unified simple form, upon which the orbital-selective physics adds on an essential and colorful light.

**Acknowledgements**

The authors would like to thank Véronique Brouet, Thomas Devereaux, Rafael Fernandes, and Leni Bascones for enlightening discussions, and David Singh for kindly providing LDA calculations for $SrFe_2P_2$. This work was supported by the U.S. Department of Energy, Office of Science, Basic Energy Sciences, Materials Sciences and Engineering Division under contract DE-AC02-76SF00515. Stanford Synchrotron Radiation Lightsource and the Advanced Light Source are both operated by the U.S. Department of Energy, Office of Science, Office of Basic Energy Sciences.

**Competing Interests**

The authors declare no competing financial interests.

**Author Contributions**

M.Y. organized the section on the normal state. Y. Z. organized the sections on the nematic, spin-density-wave, and superconducting states. Z.X.S. and D.H.L. advised and oversaw the overall structure of the manuscript. All authors contributed to the writing of the manuscript.

**Data availability**

The data sets used for the normal state properties were taken from available literature and the references are all provided in Table 1.

Coexistence of Spin-Density-Wave and Superconductor Phases in Single Crystalline Sr$_{1-x}$K$_x$Fe$_2$As$_2$. *Phys. Rev. Lett.* **102**, 127003 (2009).

**Figure Legends**

**Figure 1: General phase diagram and electronic structure of iron-based superconductors.** (a) General phase diagram of iron-based superconductors. The inset shows the lattice structure of the common Fe-As plane. (b) Lattice and magnetic structure in the symmetry breaking states of iron-based superconductors. (c) LDA calculation of the band dispersion along the Γ-M direction in NaFeAs (adapted from Ref. 30, previously published under CC BY-NC-SA license). Right panel defines the different orbitals and their color-coding used throughout the paper. (d) Illustration of the general Fermi surface topology of an iron-based superconductor.

**Figure 2: Trends of orbital-dependent electronic correlations in iron-based superconductors.** (a)-(c) Second energy derivatives of the photoemission spectra images taken along the Γ-M direction in $SrFe_2P_2$ (measured at 10K at 42.5eV under *s* polarization), NaFeAs (adapted from Ref. 30, previously published under CC BY-NC-SA license), and $FeSe_{0.44}Te_{0.56}$ (adapted from Ref. 16, previously published under CC-BY license), respectively, showing the progression from a phosphide to an arsenide to a chalcogenide. The solid lines and numbers are the band dispersions and corresponding band renormalization factors. The renormalization factors are determined by the ratio between the LDA calculated bandwidth and the experimental one. The experimental bandwidths are illustrated by the color bars to the right side of the images. (d) $d_{yz}$ band renormalization factor as a function of Fe-Pn/Ch bond length (defined by the inset). To remove the electron filling effect, we only include compounds that consist of 6 electrons per

Fe. (e) $d_{yz}$ band renormalization factor as a function of electron filling in Li(Fe,Co)As and (Ba,K)(Fe,Co,Ni)$_2$As$_2$. The phase diagram of Co and K doped BaFe$_2$As$_2$ is also plotted in the background where pink (green) region indicates the superconducting (SDW) phase. (f) $d_{yz}$ band renormalization factor for all available compounds as functions of bond length and electron filling. The color of the markers in (d) and (f) indicates the $d_{yz}$ renormalization factor as shown in the color bar. (g) The renormalization ratio between $d_{xy}$ and $d_{yz}$ bands as a function of the $d_{yz}$ renormalization factor. (h) The renormalization ratio between $d_{xy}$ and $d_{yz}$ bands as a function of bond angle (defined by the inset). (i) Overall plot of the renormalization ratio between $d_{xy}$ and $d_{yz}$ bands for all available compounds as functions of bond angle and the $d_{yz}$ renormalization factor. The color of the markers in (g)-(i) indicates the renormalization ratio between $d_{xy}$ and $d_{yz}$ as shown in the color bar. (See Table 1 for references used in generating (d)-(i).) (Panel a contains unpublished data.)

**Figure 3: Renormalization factors against lattice constants.** The $d_{yz}$ renormalization factors against lattice constants (a) *a* and (b) *c* for undoped compounds. The same for the $d_{xy}$ renormalization factors against lattice constants (c) *a* and (d) *c* for undoped compounds. The lattice constant *c* is adjusted to be per Fe layer for comparison among different structural families by dividing by two for the 122 structures. (See Table 1 for references used in generating all panels.)

**Figure 4: Orbital selective Mott phase in iron-based superconductors.** (a) $d_{xy}$ band renormalization factor as a function of bond length (defined by the inset). The color of the markers indicates the $d_{xy}$ renormalization factor as shown in the color bar. (See Table 1 for references used.) (b) The spectral weights of $d_{xy}$ and $d_{yz}$ bands as a function of temperature showing the disappearance of the $d_{xy}$ spectral weight while that of the $d_{yz}$ orbital remains finite (adapted from previously published Fig. 5h in ref. 16 under CC-BY license). (c) Theoretically calculated phase diagram of the orbital selective Mott phase (OSMP) as a function of temperature, Coulomb repulsion U, and electron filling. The blue shading indicates the spectral weight of the $d_{xy}$ orbital, and the solid blue lines indicate where the $d_{xy}$ spectral weight drops to zero, marking the boundary of the OSMP. The red line indicates the boundary of a Mott insulating (MI) phase for integer filling (adapted from previously published Fig. 6 of ref. 16 under CC_BY license).

**Figure 5: The band splitting and nematic state in iron-based superconductors.** (a) Illustration of the ARPES measurement on a twinned sample, where the beam spot encompasses multiple orthogonal domains. (b) The second derivative photoemission spectra images taken along the $a_O$ and $b_O$ directions in detwinned NaFeAs (adapted from Fig. 6 of ref. 31). (c) The second derivative photoemission spectra image taken along the Γ-M direction at 140 and 70 K in 35-layer FeSe film grown on $SrTiO_3$ (adapted from Fig. 2 of ref. 36). (d) The absolutely band

splitting energy as a function of momentum along Γ-M (adapted from Fig. 3a of ref. 36). (e) Illustration of the momentum-dependent orbital anisotropy along Γ-M (adapted from Fig. 3c of ref. 36). (f) Temperature-dependent second derivative photoemission spectra images across the M point of 35-layer FeSe film grown on $SrTiO_3$ (adapted from Fig. 4b of ref. 36). (g) Illustration of the band evolution when the system goes through a nematic transition, including the splitting of $d_{xz}/d_{yz}$ and of $d_{xy}$ (adapted from Fig. 5 of ref. 36).

**Figure 6: Fermi surface reconstruction in the magnetic state in iron-based superconductors.** (a) Illustration of the Fermi surface folding in the magnetic state in the 1-Fe BZ where folding occurs along the AFM direction to induce small Fermi surfaces. (b) and (c) Illustration and second derivative plots of the folded $\Gamma/M_x$ point along the AFM and FM directions. (d) and (e) Illustration and second derivative plots of the folded $M_y/\Gamma'$ point along the AFM and FM directions. Data were taken on detwinned $BaFe_2As_2$ within the SDW state, with photon energies and in-plane polarizations labeled in white and red, respectively. (Panels b-e contain unpublished data.)

**Tables**

| Label | Compound | filling | $a$ (Å) | $c$ (Å) | $d_{MX}$ (Å) | $\theta_2$ (°) | $W_{yz}$ (eV) | $W_{yz}^{DFT}$ (eV) | $R_{yz}$ | $R_{xy}$ | $R_{xy}/R_{yz}$ | Refs. |
|---|---|---|---|---|---|---|---|---|---|---|---|---|
| BFA | BaFe$_2$As$_2$ | 6 | 3.9622 | 13.001 | 2.3980 | 111.41 | 0.120 | 0.405 | 3.50 | 3.51 | 1.00 | 86, 7 |
| BC1.6 | Ba(Fe$_{0.984}$Co$_{0.016}$)$_2$As$_2$ | 6.016 | 3.9621 | 12.9957 | 2.3985 | 111.375 | 0.120 | 0.404 | 3.37 | | | 86,87 |
| BC2.5 | Ba(Fe$_{0.975}$Co$_{0.025}$)$_2$As$_2$ | 6.025 | 3.9621 | 12.9927 | 2.3988 | 111.355 | 0.116 | 0.404 | 3.48 | | | 86, 87 |
| BC3.5 | Ba(Fe$_{0.965}$Co$_{0.035}$)$_2$As$_2$ | 6.035 | 3.9621 | 12.9894 | 2.3991 | 111.333 | 0.126 | 0.403 | 3.20 | | | 86,87 |
| BC4.5 | Ba(Fe$_{0.955}$Co$_{0.045}$)$_2$As$_2$ | 6.045 | 3.9620 | 12.9861 | 2.3994 | 111.311 | 0.128 | 0.403 | 3.14 | | | 86,87 |
| BC5.6 | Ba(Fe$_{0.944}$Co$_{0.056}$)$_2$As$_2$ | 6.056 | 3.9620 | 12.9825 | 2.3997 | 111.286 | 0.132 | 0.402 | 3.05 | | | 86,87 |
| BC7 | Ba(Fe$_{0.93}$Co$_{0.07}$)$_2$As$_2$ | 6.07 | 3.9619 | 12.9778 | 2.4001 | 111.255 | 0.130 | 0.401 | 3.09 | | | 86,87 |
| BC7.5 | Ba(Fe$_{0.925}$Co$_{0.075}$)$_2$As$_2$ | 6.075 | 3.9619 | 12.9762 | 2.4003 | 111.244 | 0.132 | 0.401 | 3.04 | 2.76 | 0.91 | 86,87,88 |
| BC15 | Ba(Fe$_{0.85}$Co$_{0.15}$)$_2$As$_2$ | 6.15 | 3.9616 | 12.9513 | 2.4025 | 111.079 | 0.141 | 0.397 | 2.82 | | | 86,87,89 |
| BCA | BaCo$_2$As$_2$ | 7 | 3.958 | 12.67 | 2.428 | 109.2 | 0.250 | 0.350 | 1.40 | 0.980 | 0.70 | 87 |
| BK4.2 | Ba$_{0.958}$K$_{0.042}$Fe$_2$As$_2$ | 5.979 | 3.9572 | 13.0362 | 2.3978 | 111.213 | 0.117 | 0.412 | 3.52 | | | 86,90,84 |
| BK5.5 | Ba$_{0.945}$K$_{0.055}$Fe$_2$As$_2$ | 5.9725 | 3.9556 | 13.047 | 2.3978 | 111.152 | 0.117 | 0.415 | 3.54 | | | 86,90,84 |
| BK15 | Ba$_{0.85}$K$_{0.15}$Fe$_2$As$_2$ | 5.925 | 3.9442 | 13.1265 | 2.3974 | 110.707 | 0.103 | 0.431 | 4.19 | | | 86,90,84 |
| BK40 | Ba$_{0.6}$K$_{0.4}$Fe$_2$As$_2$ | 5.8 | 3.9141 | 13.3358 | 2.3964 | 109.534 | 0.102 | 0.476 | 4.66 | | | 86,90,91 |
| BK70 | Ba$_{0.3}$K$_{0.7}$Fe$_2$As$_2$ | 5.65 | 3.8781 | 13.5869 | 2.3952 | 108.127 | 0.085 | 0.528 | 6.22 | | | 86,90,92 |
| BM8 | Ba(Fe$_{0.92}$Mn$_{0.08}$)$_2$As$_2$ | 5.92 | 3.9781 | 13.0388 | 2.4117 | 111.172 | 0.111 | 0.355 | 3.58 | | | 86,93,94 |
| BNA | BaNi$_2$As$_2$ | 8 | 4.142 | 11.65 | 2.405 | 118.9 | | | 1.66 | | | 95 |
| BNP | BaNi$_2$P$_2$ | 8 | 3.947 | 11.820 | 2.260 | 121.71 | 0.544 | 0.645 | 1.18 | | | 96,97 |
| BP15 | BaFe$_2$(As$_{0.85}$P$_{0.15}$)$_2$ | 6 | 3.9444 | 12.914 | 2.3772 | 112.157 | 0.125 | 0.427 | 3.41 | 3.25 | 0.95 | 86,7 |
| BP20 | BaFe$_2$(As$_{0.8}$P$_{0.2}$)$_2$ | 6 | 3.9385 | 12.885 | 2.3702 | 112.406 | 0.133 | 0.434 | 3.26 | 2.68 | 0.82 | 86,7 |
| BP30 | BaFe$_2$(As$_{0.7}$P$_{0.3}$)$_2$ | 6 | 3.9266 | 12.827 | 2.3563 | 112.904 | 0.145 | 0.449 | 3.09 | 2.11 | 0.68 | 86,7 |
| BP50 | BaFe$_2$(As$_{0.5}$P$_{0.5}$)$_2$ | 6 | 3.9029 | 12.712 | 2.3285 | 113.9 | 0.171 | 0.478 | 2.79 | 1.81 | 0.65 | 86,7 |
| CFA | CaFe$_2$As$_2$ | 6 | 3.872 | 11.730 | 2.370 | 109.5 | 0.109 | 0.338 | 3.10 | | | 98,99 |
| CFAcT | CaFe$_2$As$_2$-cT | 6 | 3.978 | 5.304 | 2.34 | 116.4 | 0.098 | 0.281 | 2.88 | | | 100,98,99 |
| CFP | CaFe$_2$P$_2$ | 6 | 3.855 | 9.985 | 2.240 | 118.74 | 0.278 | 0.438 | 1.57 | 1.50 | 0.96 | 101 |
| EFA | EuFe$_2$As$_2$ | 6 | 3.9062 | 12.1247 | 2.382 | 110.1 | 0.098 | 0.315 | 3.22 | | | 102,103 |
| FS | FeSe | 6 | 3.774 | 5.523 | 2.395 | 104.02 | 0.140 | 0.592 | 4.23 | | | 104,32 |
| FTS20 | FeTe$_{0.80}$Se$_{0.20}$ | 6 | 3.8156 | 6.123 | 2.5614 | 96.484 | 0.089 | 0.483 | 5.46 | | | 104,105 |
| FTS28 | FeTe$_{0.72}$Se$_{0.28}$ | 6 | 3.8114 | 6.063 | 2.545 | 97.238 | 0.099 | 0.494 | 5.00 | 22.13 | 4.42 | 104,106 |
| FTS30 | FeTe$_{0.70}$Se$_{0.30}$ | 6 | 3.8104 | 6.048 | 2.5406 | 97.426 | 0.100 | 0.497 | 4.97 | | | 104,105 |
| FTS35 | FeTe$_{0.65}$Se$_{0.35}$ | 6 | 3.8078 | 6.0105 | 2.5302 | 97.897 | 0.102 | 0.503 | 4.92 | 18.81 | 3.82 | 104,106 |
| FTS40 | FeTe$_{0.60}$Se$_{0.40}$ | 6 | 3.8052 | 5.973 | 2.5198 | 98.368 | 0.110 | 0.510 | 4.64 | | | 104,105 |
| FTS45 | FeTe$_{0.55}$Se$_{0.45}$ | 6 | 3.8026 | 5.9355 | 2.5094 | 98.839 | 0.114 | 0.517 | 4.53 | 16.46 | 3.64 | 104,106 |
| FTS59 | FeTe$_{0.41}$Se$_{0.59}$ | 6 | 3.7953 | 5.8305 | 2.4803 | 100.158 | 0.125 | 0.536 | 4.30 | 14.86 | 3.45 | 104,106 |
| KFA | KFe$_2$As$_2$ | 5.5 | 3.842 | 13.838 | 2.394 | 106.72 | 0.079 | 0.581 | 7.32 | | | 90,107 |
| KFS | KFe$_2$Se$_2$ | 6.16 | 3.9136 | 14.0367 | 2.4406 | 106.6 | 0.120 | 0.375 | 3.13 | | | 108,15 |
| LFA | LiFeAs | 6 | 3.7914 | 6.364 | 2.421 | 103.11 | 0.111 | 0.400 | 3.60 | 4.24 | 1.18 | 109,7 |
| LFOHFS | Li$_{0.8}$Fe$_{0.2}$OHFeSe | 6.1 | 3.7705 | 9.215 | 2.3941 | 103.90 | 0.118 | 0.442 | 3.75 | 7.31 | 1.95 | 110,111 |
| LFP | LaFe$_2$P$_2$ | 6.5 | 3.841 | 10.982 | 2.242 | 117.841 | 0.200 | 0.309 | 1.54 | 1.50 | 0.97 | 101 |
| LC3 | LiFe$_{0.968}$Co$_{0.032}$As | 6.032 | 3.7903 | 6.3578 | 2.4196 | 103.123 | 0.121 | 0.392 | 3.24 | | | 112,113,7 |
| LC9 | LiFe$_{0.907}$Co$_{0.093}$As | 6.093 | 3.7883 | 6.3460 | 2.4179 | 103.148 | 0.140 | 0.384 | 2.75 | | | 112,113,7 |
| LC12 | LiFe$_{0.877}$Co$_{0.123}$As | 6.123 | 3.7873 | 6.3401 | 2.4170 | 103.16 | 0.153 | 0.380 | 2.49 | | | 112,113,7 |
| LC17 | LiFe$_{0.83}$Co$_{0.17}$As | 6.17 | 3.7857 | 6.3310 | 2.4157 | 103.179 | 0.175 | 0.374 | 2.41 | 2.23 | 1.04 | 112,113,7 |
| LC30 | LiFe$_{0.7}$Co$_{0.3}$As | 6.3 | 3.7814 | 6.3058 | 2.4120 | 103.232 | 0.239 | 0.357 | 1.49 | 1.60 | 1.07 | 112,113,7 |
| LFPO | LaFePO | 6 | 3.941 | 8.507 | 2.280 | 119.62 | 0.236 | 0.558 | 2.37 | 2.62 | 1.11 | 6 |
| NFA | NaFeAs | 6 | 3.9448 | 6.9968 | 2.4282 | 108.64 | 0.128 | 0.516 | 4.03 | 6.98 | 1.73 | 114, 7,30 |
| SFA | SrFe$_2$As$_2$ | 6 | 3.9243 | 12.3644 | 2.388 | 110.5 | 0.107 | 0.354 | 3.32 | | | 115,116 |
| SFP | SrFe$_2$P$_2$ | 6 | 3.825 | 11.612 | 2.251 | 116.4 | 0.264 | 0.499 | 1.89 | 1.84 | 0.97 | 115 |
| SK20 | Sr$_{0.8}$K$_{0.2}$Fe$_2$As$_2$ | 5.9 | 3.9115 | 12.6214 | 2.3892 | 109.744 | 0.091 | 0.400 | 4.40 | | | 115,90,116 |

**Table 1**: **Material details from literature used for the normal state.** $d_{MX}$ is the bond length between the transition element ($M$ = Fe, Co, Ni, Mn) and pnictogen/chalcogen ($X$ = P, As, Se, Te). $\theta_2$ is the two-fold $X-M-X$ bond angle within an $M$-centered $M\times 4$ tetrahedron. $W_{yz}$ and $W_{yz}^{DFT}$ are the d$_{yz}$ bandwidth of the data and DFT calculations, respectively, taken as the energy range from the band top at the Γ point and the band bottom somewhere between the Γ point and the M point. If the band top is above $E_F$ and hence not observable, a parabolic fitting of the observable parts of the bands around Γ is used to determine the band top position. $R_{yz}$, the d$_{yz}$ band

renormalization, is determined by the ratio of $W_{yz}^{DFT}$ and $W_{yz}$. $R_{xy}$, the $d_{xy}$ band renormalization, is taken as the ratio of the fitted band slope of the most visible section of the LDA-calculated and measured hole-dispersion near Γ. Structural parameters and $W_{yz}^{DFT}$ for intermediate dopings of a material series are linear interpolations of the end members. The structural parameters are taken from Ref. 3 unless other references are given. All measurement info is taken from the normal state spectra to avoid complications from the spin density wave or orthorhombic phase, and taken at $k_z = 0$ wherever possible. Where references are not given, the data come from unpublished data of the authors. Only compounds with sufficiently complete information including ARPES data, LDA calculation, and relevant information such as structural parameter and/or electron filling are included in each plot. Compounds are ordered alphabetically by label.

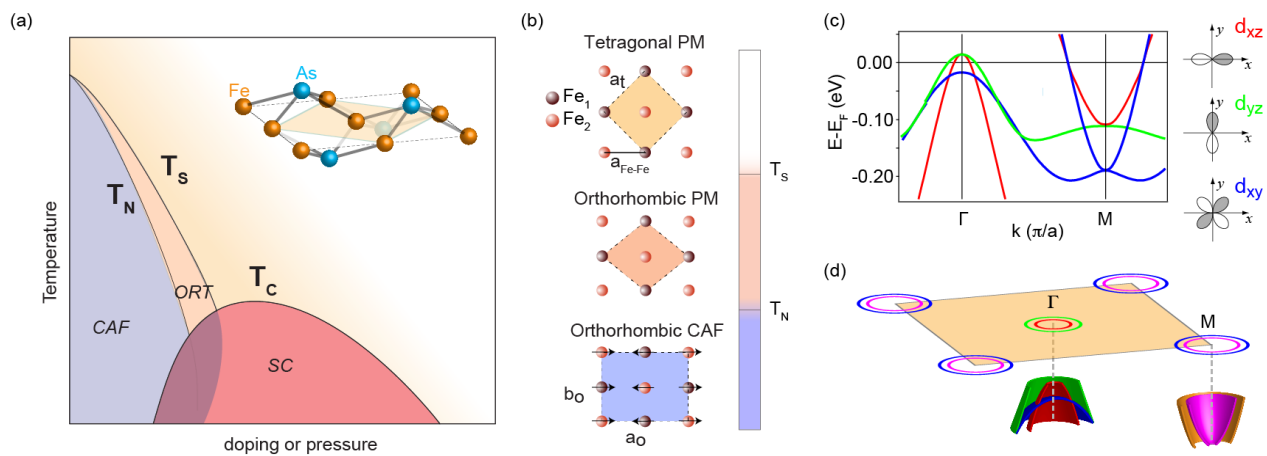

**Figure 1**

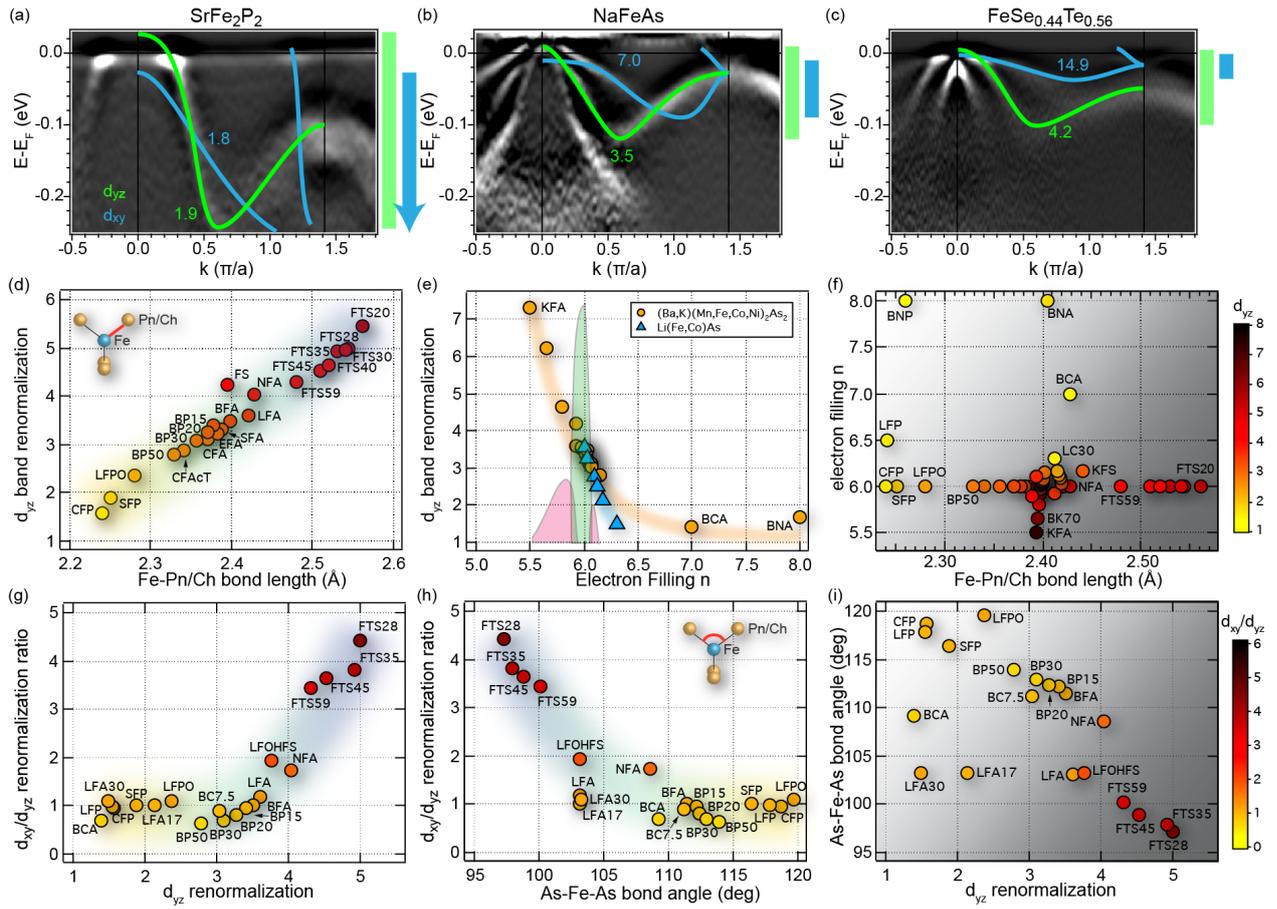

**Figure 2**

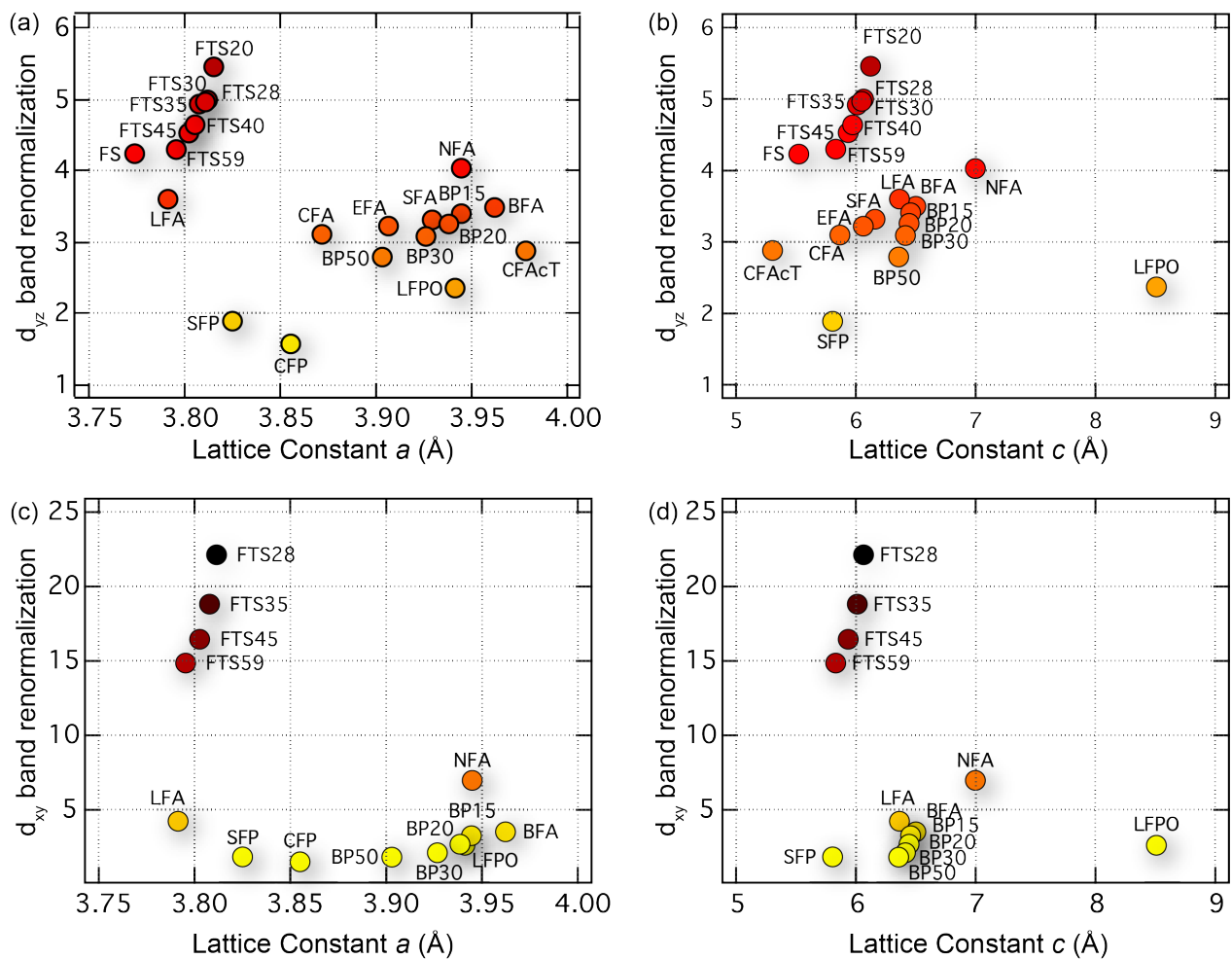

**Figure 3**

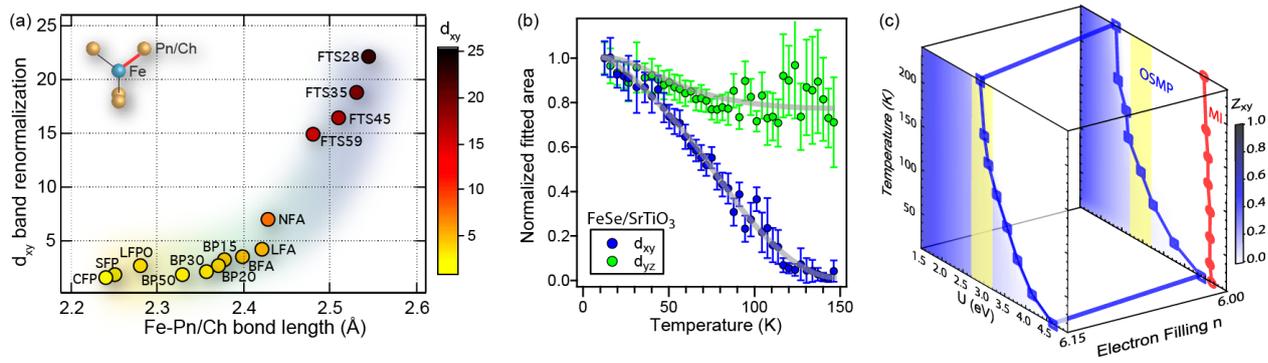

**Figure 4**

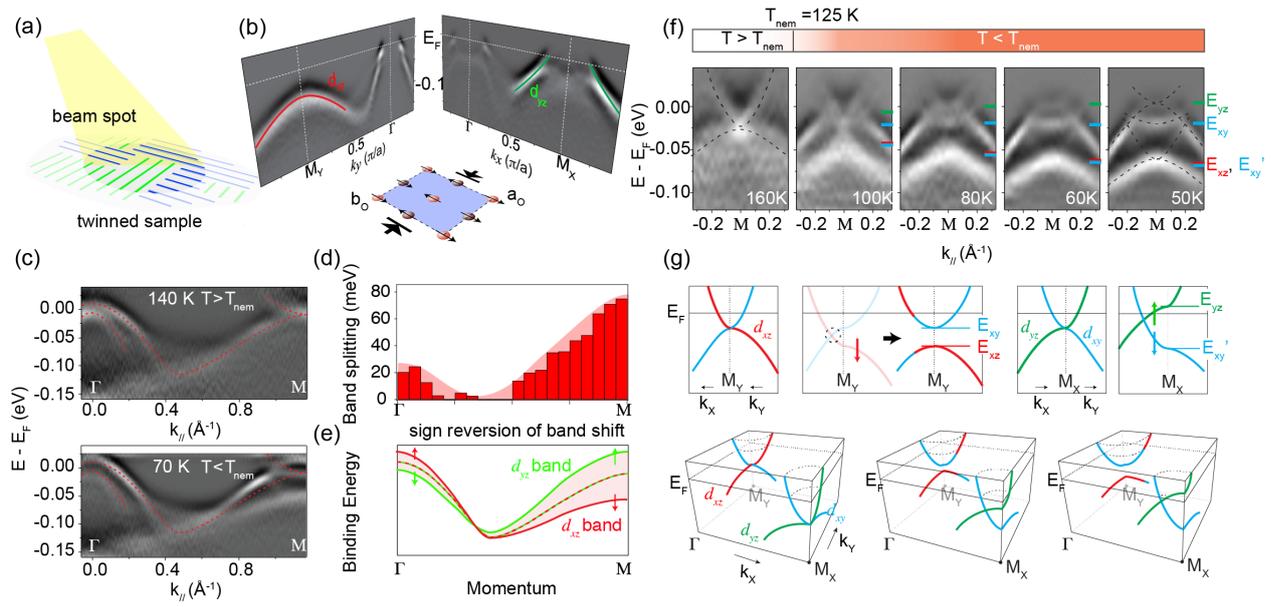

**Figure 5**

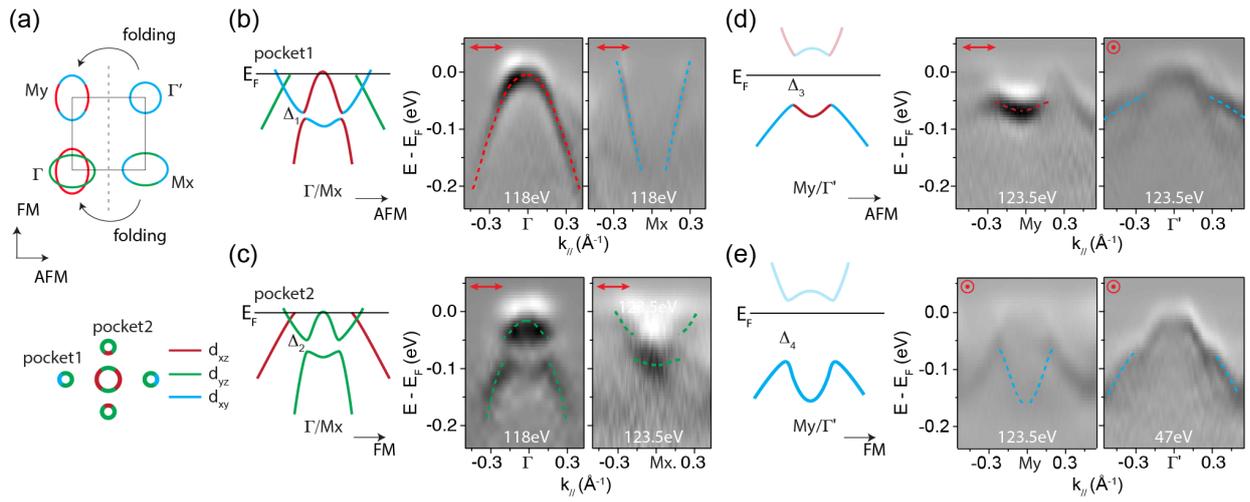

**Figure 6**

50